\documentclass
[prd,aps,twocolumn,showpacs,amsmath,amssymb,nofootinbib]{revtex4}

\usepackage{graphicx}  
\usepackage{bm}        
\usepackage{paralist}  
\usepackage{amsthm}    

\theoremstyle{plain}
\newtheorem{thm}{Theorem}
\newtheorem{pro}[thm]{Proposition}

\theoremstyle{definition}

\theoremstyle{remark}

\newcommand{\ul}[1]{\underline{#1}}

\newcommand{\sss}{\scriptscriptstyle}
\newcommand{\ind}[1]{{\sss \text{#1}}}        
\DeclareMathAlphabet{\bi}{OML}{cmm}{b}{it}    
\newcommand{\br}[1]{\mathbf{#1}}              

\newcommand{\abs}[1]{\lvert#1\rvert}          
\newcommand{\norm}[1]{\lVert#1\rVert}         
\newcommand{\sprod}[2]{\langle#1,#2\rangle}   

\DeclareMathOperator{\LC}{\bm{\nabla}}        
\DeclareMathOperator{\Grad}{\LC\!}            
\DeclareMathOperator{\Div}{\br{div}}          
\DeclareMathOperator{\Riem}{\br{Riem}}        
\DeclareMathOperator{\Weyl}{\br{Weyl}}        
\DeclareMathOperator{\Ric}{\br{Ric}}          
\DeclareMathOperator{\Scal}{\mathrm{Scal}}    
\DeclareMathOperator{\Ein}{\br{Ein}}          
\newcommand{\bpartial}{\boldsymbol{\partial}} 
\newcommand{\bnabla}{\boldsymbol{\nabla}}     
\newcommand{\bdot}{\boldsymbol{\cdot}}
\newcommand{\ed}{\bi{d}}                      

\newcommand{\M}{\mathcal{M}}                      
\newcommand{\Stwo}{{S^2}}                         
\newcommand{\sStwo}{{\sss \Stwo}}                 
\newcommand{\gStwo}{\g_\sStwo}                    

\newcommand{\Id}{\mathbf{id}}                   
\newcommand{\g}{\bg}                            

\newcommand{\be}{\bi{e}}

\newcommand{\bg}{\bi{g}}
\newcommand{\bh}{\bi{h}}

\newcommand{\bk}{\bi{k}}
\newcommand{\bl}{\bi{l}}

\newcommand{\bq}{\bi{q}}
\newcommand{\bs}{\bi{s}}

\newcommand{\bu}{\bi{u}}
\newcommand{\bv}{\bi{v}}

\newcommand{\bP}{\bi{P}}

\newcommand{\bT}{\bi{T}}

\newcommand{\bX}{\bi{X}}
\newcommand{\bY}{\bi{Y}}

\newcommand{\bsigma}{\boldsymbol{\sigma}}

\newcommand{\Lemaitre}{Lema\^{\i}tre}

\newcommand{\gSchw}{\g^\ind{Schw}_{\sss M_0}}
\newcommand{\gcS}{\g^\ind{cS}_{\sss \Omega, M_0}}
\newcommand{\gMcV}{\g^\ind{McV}_{a,m}}
\newcommand{\buflat}{\text{\ul{$\bu$}}}
\newcommand{\bkflat}{\text{\ul{$\bk$}}}

\begin{document}


\title{On the generalization of McVittie's model\\
       for an inhomogeneity in a cosmological spacetime}

\author{Matteo Carrera}
\email{matteo.carrera@physik.uni-freiburg.de}
\affiliation{Institute of Physics, University of Freiburg, 
Hermann-Herder-Strasse~3, D-79104 Freiburg, Germany}
\author{Domenico Giulini}
\email{domenico.giulini@itp.uni-hannover.de}
\affiliation{University of Hannover,
Appelstrasse~2, D-30167 Hannover, Germany}
\altaffiliation[Also at: ]{ZARM, University of Bremen, Am Fallturm\\ 
D-28359 Bremen, Germany}

\date{August 21, 2009}

\begin{abstract}
McVittie's spacetime is a spherically symmetric solution to Einstein's 
equation with an energy-momentum tensor of a perfect fluid. It describes 
the external field of a single quasi-isolated object with vanishing 
electric charge and angular momentum in an environment that asymptotically 
tends to a Friedmann--\Lemaitre--Robertson--Walker universe. 
We critically discuss some recently proposed generalizations of this 
solution, in which radial matter accretion as well as heat 
currents are allowed. We clarify the hitherto unexplained
constraints between these two generalizing aspects as being
due to a geometric property, here called spatial Ricci-isotropy, 
which forces solutions covered by the McVittie ansatz 
to be rather special. We also clarify other aspects of 
these solutions, like whether they include geometries which 
are in the same conformal equivalence class as the exterior 
Schwarzschild solution, which leads us to contradict some
of the statements in the recent literature.        
\end{abstract}


\maketitle

\tableofcontents


\section{Introduction}
\label{sec:Introduction}
Two sets of exact solutions to Einstein's field equation of 
General Relativity are of paradigmatic importance: The first 
set describes the gravitational field of quasi-isolated objects
in an asymptotically flat space-time. Among them is the 
exterior Schwarzschild solution that describes the stationary 
gravitational field outside a spherically symmetric star or 
black hole of mass $m$ with vanishing intrinsic angular momentum 
(spin) and vanishing electric charge. (The latter two features 
being included in the three-parameter Kerr--Newman family 
of solutions.) Such asymptotically flat solutions are meant 
to apply to a region outside the central object which, on 
one hand, must be sufficiently far from the considered 
object, so as to legitimately neglect small irregularities 
of its surface and/or small deviations from perfect spherical 
symmetry. On the other hand, and more importantly, the region of 
applicability must also be sufficiently close to the considered 
object in order not to include, or come close to, other 
compact sources, or not contain too much dust-filled space 
between it and the object which would also act as disturbing 
source for the gravitational field. In particular, the 
large-distance asymptotic behavior of such solutions is 
an idealization and not meant to be strictly that of any 
object in the real world. 

On the other hand, the second set of paradigmatic solutions are 
the cosmological ones, which aim to model the behavior of 
space-time at the largest cosmological scales, without trying 
to be realistic at smaller scales. Among them is the family of 
homogeneous and isotropic Friedmann--\Lemaitre--Robertson--Walker 
(FLRW) cosmologies on  which the cosmological standard-model 
is based. 

Given that situation, the task is to combine the virtues of
both classes of solutions without the corresponding 
deficiencies. This means to find exact solutions for the 
gravitational field of a compact object `immersed' (see below) 
into an otherwise cosmological background. This would appear 
to be an easy task if the field equations were linear, for, 
in that case, one would just add the solution that describes 
the gravitational field of a compact object in an otherwise 
empty universe to the cosmological solution that corresponds 
to a homogeneous distribution of background matter. 
Here the mathematical operation of addition appears to be 
the obvious realization of what one might be tempted to 
call `simultaneous physical presence' and hence, in view of 
the individual interpretations of both solutions, the 
`immersion' of the compact object into the cosmological 
background. But this immediate interpretation in physical 
terms of a simple mathematical operation is deceptive. 
This becomes obvious in non-linear theories, like General 
Relativity (GR), where no simple mathematical operation exists 
that produces a new solution out of two old ones and where 
the very same physical question may still be asked. 

The proper requirement for a mathematical representation of the 
envisaged physical situation must, first of all, consist in 
asymptotic conditions which ensure that the sought for solution 
approximates the given (e.g. Schwarzschild) one for small 
distances and a particular cosmological one (e.g. FLRW) for 
large distances. Second, it must specify somehow the physics 
in the intermediate region. Usually this will include a 
specification of the matter components and their dynamical laws 
together with certain initial and boundary conditions. Needless
to say that this will generally result in a complex system of 
partial differential equations. Most analytic approaches 
therefore impose further simplifying assumptions that 
automatically guarantee the right asymptotic behavior and at the 
same time reduce the free functions to a manageable number. 

In this paper we will discuss a particular such approach, which is 
originally due to McVittie \cite{McVittie:1933} and which has been 
further analyzed and clarified in a series of carefully written 
papers by Nolan~\cite{Nolan:1998,Nolan:1999a,Nolan:1999b}. Our main 
motivation is that recently McVittie's solution has been severely 
criticized as not being able at all to model the envisaged 
situation\cite{Gao.etal:2008,Faraoni.etal:2008}, whereas a 
family of slightly generalized ones~\cite{Faraoni.Jacques:2007}, 
in which some restrictions concerning the motion of matter and 
the existence of heat flows is lifted, is argued to be free of 
the alleged problems. The existence of an exact solution to 
Einstein's equation that models local inhomogeneities is clearly 
of great importance, for example in estimating reliable upper 
bounds to the possible influence of global cosmological expansion
onto the dynamics and kinematics of local 
systems~\cite{Carrera.Giulini:2008a}. 

The paper is organized as follows: 
In Section\,\ref{sec:McVittieModel} we review what we call the 
\emph{McVittie model}. We discuss its metric ansatz and what its 
entails regarding the geometry of spacetime. Then we discuss the 
assumptions regarding the motion of the matter and how this, 
together with Einstein's equation, determines one of the two 
free functions in the metric ansatz as a simple function of the 
other. We interpret this condition in terms of an appropriate 
concept of local mass as saying that the object does not 
accrete mass from the ambient matter.  
In Section\,\ref{sec:SecondLookMcVittie} we take a second and closer 
look at the McVittie ansatz and note some of its characteristic 
features which, we feel, have not sufficiently carefully been 
taken into account 
in~\cite{Faraoni.Jacques:2007,Gao.etal:2008,Faraoni.etal:2008}.
In the light of these observations we then discuss in 
Section\,\ref{sec:GeneralisationsMcVittie} the attempted 
generalizations of McVittie's solution in the references 
just mentioned. We find that some of the conclusions drawn 
are indeed unwarranted.

\section{The McVittie model}
\label{sec:McVittieModel}
The characterization of the McVittie model is made through two sets 
of \emph{a\,priori} specifications. The first set concerns the metric 
(left side of Einstein's equations) and the second set the matter 
(right side of Einstein's equations). The former consists in an 
ansatz for the metric, which can formally be described as follows:
Write down the Schwarzschild metric for the mass parameter $m$ in 
isotropic coordinates, add a conformal factor $a^2(t)$ to the 
spatial part, and allow the mass parameter $m$ to depend on time. 
Hence the metric reads    
\begin{equation}\label{eq:McVittieAnsatz}
\begin{split}
\g =&\left( \frac{1-m(t)/2r}{1+m(t)/2r} \right)^2 \ed t^2\\
   -&\left( 1+\frac{m(t)}{2r} \right)^4 a^2(t)\ (\ed r^2 + r^2 \gStwo)\,,
\end{split}
\end{equation}
where $\gStwo=\ed\theta^2+\sin^2\theta\ed\varphi^2$ is the standard metric 
on the unit 2-sphere. Here we restricted attention to the asymptotically 
\emph{spatially flat} (i.e. $k=0$) FLRW metric, which is compatible with 
current cosmological data~\cite{Komatsu.etal:2008}. For simplicity we shall 
refer to (\ref{eq:McVittieAnsatz}) simply as \emph{McVittie's ansatz}, 
though this is not quite correct since McVittie started from a general
spherically symmetric form and arrived at (\ref{eq:McVittieAnsatz}) 
with $m(t)a(t)=\mathrm{const.}$ after imposing a condition
that he interpreted as condition for no matter-infall. The ansatz 
(\ref{eq:McVittieAnsatz}) is obviously spherically symmetric with 
the spheres of constant radius $r$ being the orbits of the 
rotation group.%
\footnote{``Spherical symmetry'' of a spacetime means the following: 
There exists an action of the group $SO(3)$ on spacetime by isometries, 
which is such that the orbits are either two-dimensional and spacelike 
or fixed points. The ``spheres'' implicitly referred to in this term 
correspond to the two-dimensional orbits, even though they might in 
principle also be two-dimensional real projective spaces. In the cases 
we discuss here they will be 2-spheres.} 
In the next section we will discuss in more detail the geometric 
implications of this ansatz, independent of whether Einstein's equation
holds.  

As already discussed in the introduction, the model here is meant to 
interpolate between the spherically symmetric gravitational field of 
a compact object and the environment. It is not to be taken too 
seriously in the region very close to the central object, where 
the basic assumptions on the behavior of matter definitely turn 
unphysical. However, as discussed in~\cite{Carrera.Giulini:2008a}, 
at radii much larger than (in geometric units) the central mass 
(to be defined below) the $k=0$ McVittie solution seems to provide 
a viable approximation for the envisaged situation. 

The second set of specifications, concerning the matter, is as follows: 
The matter is a perfect fluid with density $\varrho$ and isotropic 
pressure $p$. Hence its energy-momentum tensor is given by%
\footnote{Here and in what follows we denote the metric-dual (1-form) of 
a vector $\bu$ by underlining it, that is, $\buflat:=\bg(\bu,\,\cdot\,)$ 
is the 1-form metric-dual to the vector $\bu$. In local coordinates we 
have $\bu=u^\mu\bpartial_\mu$ and $\buflat=u_\mu \ed x^\mu$, where 
$u_\mu:=g_{\mu\nu}u^\nu$.} 
\begin{equation}\label{er:EMTensorForMcVittie}
  \bT = \varrho\,\ul\bu\otimes\ul\bu +p\,(\ul\bu\otimes\ul\bu-\bg)\,. 
\end{equation}
Furthermore, and this is where the two sets of specifications make 
contact, the motion of the matter (i.e. its four-velocity field)
is given by  
\begin{equation}\label{eq:u-McVittie}
  \bu = \be_0 \,,
\end{equation}
where $\be_0$ is the normalization of $\bpartial/\bpartial t$ 
(compare~(\ref{eq:DefOrthnFrame})). 
Finally, the explicit cosmological constant on the left-hand side of 
Einstein's equation is assumed to be zero, which implies no loss of 
generality, since a non-zero cosmological constant can always be 
regarded as special part of the matter's energy-momentum tensor 
(compare~\ref{sec:GeneralisationsMcVittie}).
No further assumptions are made. In particular, an equation of state, 
like $p=p(\varrho)$, is \emph{not} assumed. The reason for this 
will become clear soon. Later generalizations will mainly concern 
(\ref{er:EMTensorForMcVittie}) and (\ref{eq:u-McVittie}). 

The Einstein equation\footnote{We speak of ``the Einstein equation'' 
in the singular since we think of it as a single tensor equation,
which only upon introducing a coordinate system decomposes in 
many scalar equations.} now links the specifications of geometry with that 
of matter. It is equivalent to the following three relations between 
the four functions $m(t), a(t), \varrho(t,r)$, and $p(t,r)$: 
\begin{subequations}\label{eq:McV-Einstein}
\begin{alignat}{2}
\label{eq:McV-Einstein-1}
  &(a \, m)\!\dot{\phantom{I}}\! &&\,=\, 0 \,,\\
\label{eq:McV-Einstein-2}
  &8\pi \varrho    &&\,=\, 3 \left( \frac{\dot a}{a} \right)^2 \,,\\
\label{eq:McV-Einstein-3}
  &8\pi p          &&\,=\, - 3 \left( \frac{\dot a}{a} \right)^2
  - 2 \left( \frac{\dot{a}}{a} \right) \!\!\!\!\dot{\phantom{\frac{I}{I}}}
  \left( \frac{1+m/2r}{1-m/2r} \right) \,.
\end{alignat}
\end{subequations}
Note that here Einstein's equation has only three independent components 
(as opposed to four for a general spherically symmetric metric), which is 
a consequence of the fact that the Einstein tensor for the McVittie 
ansatz~(\ref{eq:McVittieAnsatz}) is spatially isotropic. This will be 
discussed in more detail in the next section.   

Equation~(\ref{eq:McV-Einstein-1}) can be immediately integrated:
\begin{equation}\label{eq:mIntegration}
  m(t) = \frac{m_0}{a(t)} \,,
\end{equation}
where $m_0$ is an integration constant. Below we will show that this 
integration constant is to be interpreted as the mass of the central
body. 

Clearly the system (\ref{eq:McV-Einstein}) is under-determining. 
This is expected since no equation of state has yet been imposed. 
The reason why we did not impose such a condition can now be easily 
inferred from (\ref{eq:McV-Einstein}): whereas  (\ref{eq:McV-Einstein-2}) 
implies that $\varrho$ only depends on $t$, (\ref{eq:McV-Einstein-3}) 
implies that $p$ depends on $t$ \emph{and} $r$ iff $(\dot a/a)\dot{}\ne 0$. 
Hence a non-trivial relation $p=p(\varrho)$ is simply incompatible with the 
assumptions made so far. 
The only possible ways to specify $p$ are $p=0$ or $\varrho+p=0$. 
In the first case (\ref{eq:McV-Einstein-3}) implies that $\dot a/a=0$
if $m_0\ne 0$ (since then the second term on the right-hand side is
$r$ dependent, whereas the first is not, so that both must vanish
separately), which corresponds to the exterior Schwarzschild 
solution, or $a(t)\propto t^{2/3}$ if $m_0=0$, which leads to the 
flat FLRW solution with dust. In the second case the fluid just acts 
like a cosmological constant $\Lambda=8\pi\varrho$ (using the equation 
of state $\varrho+p=0$ in $\Div\bT=0$ it implies $\ed p=0$ and this, 
in turn, using again the equation of state, implies $\ed \varrho=0$) so 
that this case reduces to the Schwarzschild--de\,Sitter solution. 
To see this explicitly, notice first that 
(\ref{eq:McV-Einstein-2},\ref{eq:McV-Einstein-3}) imply the constancy 
of $H=\dot a/a=\sqrt{\Lambda/3}$ and hence one has 
$a(t)=a_0\exp\bigl(t\,\sqrt{\Lambda/3}\bigr)$. With such a scale-factor 
the McVittie metric~(\ref{eq:McVittieAnsatz}) with~(\ref{eq:mIntegration}) 
turns into the Schwarzschild--de\,Sitter metric in spatially isotropic 
coordinates. The explicit formulae for the coordinate transformation 
which brings the latter in the familiar form can be found in Section\,5 
of~\cite{Robertson:1928} and also in Section\,7 of~\cite{Klioner.Soffel:2005}. 
Finally, note from (\ref{eq:McV-Einstein-1}) that constancy of one of the 
functions $m$ and $a$ implies constancy of the other. In this case 
(\ref{eq:McV-Einstein-2},\ref{eq:McV-Einstein-3}) imply $p=\varrho=0$, so 
that we are dealing with the exterior Schwarzschild spacetime. 

A specific McVittie solution can be obtained by \emph{choosing} a function 
$a(t)$, corresponding to the scale function of the FLRW spacetime which 
the McVittie model is required to approach at spatial infinity, and the 
constant $m_0$, corresponding to the `central 
mass'. Relations~(\ref{eq:McV-Einstein-2},\ref{eq:McV-Einstein-3}), 
and~(\ref{eq:mIntegration}) are then used to determine $\varrho$, 
$p$, and $m$, respectively. Clearly this `poor man's way' to solve 
Einstein's equation holds the danger of arriving at unrealistic 
spacetime dependent relations between $\varrho$ and $p$. This must 
be kept in mind when proceeding in this fashion. For further 
discussion of this point we refer to~\cite{Nolan:1998,Nolan:1999a}. 

As will be discussed in more detail in 
Section\,\ref{sec: MisnerSharpEnergy} below, in the spherically-symmetric
case the concept of local mass (or energy) is well captured by the 
Misner--Sharp (MS) energy~\cite{Misner.Sharp:1964}, whose purely 
geometric definition in terms of Riemannian curvature allows to 
decompose it into a sum of two terms, one of which comes from the 
Ricci- the other from the Weyl curvature. It is the latter which may 
be identified with the gravitational mass of the central object. 
Applied to (\ref{eq:McVittieAnsatz}), the Weyl contribution to the 
MS energy can be written in the following form, also taking into 
account~(\ref{eq:mIntegration}), 
\begin{subequations}\label{eq:McV-MSE-wEeq}
\begin{alignat}{3}
  &E_\ind{R} &&= \frac{4\pi}{3}R^3 \varrho \,, 
  \label{eq:McV-MSE-Ricci-wEeq} \\
  &E_\ind{W} &&= m_0 \,.
  \label{eq:McV-MSE-Weyl-wEeq}
\end{alignat}
\end{subequations}
The constancy of $E_\ind{W}$ is then interpreted as saying that 
no energy is accreted from the ambient matter onto the central 
object.  

We now briefly discuss the basic properties of the motion of cosmological 
matter. Being spherically symmetric, the velocity field $\bu$ specified 
in~(\ref{eq:u-McVittie}) is automatically vorticity free. The last property 
is manifest from its hypersurface orthogonality, which is immediate 
from~(\ref{eq:McVittieAnsatz}). Moreover, $\bu$ is also shear free. 
This, too, can be immediately read off~(\ref{eq:McVittieAnsatz}) once 
one takes into account the following result, whose proof we sketch in 
Appendix\,\ref{sec:ProofShearFree}: A spherically symmetric normalized 
timelike vector field $\bu$ in a spherically symmetric spacetime 
$(\M,\g)$ is shear free iff its corresponding spatial metric, that is,
the metric $\g$ restricted to the subbundle 
$\bu^\perp:=\{\bv\in T\M\mid \g(\bv,\bu)=0\}$, is conformally flat. 
The metric (\ref{eq:McVittieAnsatz}) obviously is spatially conformally 
flat with respect to the choice~(\ref{eq:u-McVittie}) made here. 
Moreover, the expansion (divergence) of $\bu$ is 
\begin{equation}\label{eq:expansion-McVittie}
  \theta = 3 H\,, 
\end{equation}
where $H:=\dot a/a$, just as in the FLRW case. In particular, the expansion 
of the cosmological fluid is \emph{homogeneous in space}. Exactly as in the 
FLRW case is also the expression for the variation of the areal radius 
along the integral lines of $\bu$ (that is the velocity of cosmological 
matter measured in terms of its proper time and the areal radius): 
\begin{equation}\label{eq:Hubble-law-McVittie}
  \bu(R) = HR \,,
\end{equation}
which is nothing but Hubble's law. Recall that for a spherically symmetric 
spacetime the \emph{areal radius}, denoted here by $R$, is the function 
defined by $R(p):=\sqrt{\mathcal{A}(p)/4\pi}$, where $\mathcal{A}(p)$ is 
the proper area of the 2-dimensional $SO(3)$-orbit through the 
point $p$. For the McVittie spacetime the areal radius is given explicitly 
in~(\ref{eq:McVittieArealRadius}). 
The acceleration of $\bu$, which in contrast to the FLRW case 
does not vanish here, is given by 
\begin{equation}\label{eq:accel-McVittie}
  \bnabla_\bu\bu = \frac{m_0}{R^2} \left(\frac{1+m/2r}{1-m/2r}\right) \be_1\,.
\end{equation}
Here $\be_1$ is the normalized vector field in radial direction as 
defined in~(\ref{eq:DefOrthnFrame}). In leading order in $m_0/R$ this 
corresponds to the acceleration of the observers moving along the timelike 
Killing field in Schwarzschild spacetime. 
%
%

It is also important to note that the central gravitational mass in 
McVittie's spacetime may be modeled by a shear-free perfect-fluid star 
of positive homogeneous energy density~\cite{Nolan:1992}. The matching 
is performed along a world-tube comoving with the cosmological fluid, 
across which the energy density jumps discontinuously. This means that 
the star's surface is comoving with the cosmological fluid and hence, 
in view of~(\ref{eq:expansion-McVittie}), that it geometrically expands 
(or contracts). This feature, however, should be merely seen as an artifact 
of the McVittie model (in which the relation~(\ref{eq:expansion-McVittie}) 
holds), rather than a general property of compact objects in any cosmological 
spacetimes. Positive pressure within the star seems to be only possible if 
$2a\ddot a+{\dot a}^2<0$ (see Eq.~(3.27) in~\cite{Nolan:1992} with 
$a=\exp(\beta/2)$), that is, for deceleration parameters $q>1/2$.

\section{Geometry of the McVittie ansatz}
\label{sec:SecondLookMcVittie}
In this section we will discuss the geometry of the metric 
(\ref{eq:McVittieAnsatz}) independent of the later restriction 
that it will have to satisfy Einstein's equation for some 
reasonable energy-momentum tensor. This means that at this
point we shall not assume any relation between the two 
functions $m(t)$ and $a(t)$, apart from the first being non
negative and the second being strictly positive.  We will 
discuss the metric's `spatial Ricci-isotropy' (a term explained 
below), its singularities and trapped regions, and also 
compute its Misner--Sharp energy decomposed into the Ricci 
and Weyl parts. We shall start, however, by answering the 
question of what the overlap is between the geometries 
represented by (\ref{eq:McVittieAnsatz}) and the conformal 
equivalence class of the exterior Schwarzschild geometry.

\subsection{Relation to conformal Schwarzschild class}
\label{sec:ConfSchwClass}
This question is an obvious one in view of the way in which 
(\ref{eq:McVittieAnsatz}) is obtained from the exterior Schwarzschild 
metric. It is clear that for $m=m_0=\mathrm{const.}$ the metric 
(\ref{eq:McVittieAnsatz}) is conformally equivalent to the exterior 
Schwarzschild metric, since upon using a new time coordinate $T$ with 
$dT=dt/a(t)$ we can pull out $a^2(t)$ as a common conformal factor. 
The following proposition, whose proof we shall give in 
Appendix~\ref{sec:Intersection-McV-cS}, states that a constant 
$m$ is in fact also necessary condition:
\begin{pro}
\label{pro:Intersection-McV-cS}
Let $\mathcal{S}_\ind{\emph{McV}}$ denote the set of metrics 
in the form of the McVittie ansatz~(\ref{eq:McVittieAnsatz}) 
(parametrized by the two positive functions $a$ and $m$) and 
$\mathcal{S}_\ind{\emph{cS}}$ the set of metrics conformally 
equivalent to an exterior Schwarzschild metric (parametrized 
by a positive conformal factor and a constant positive 
Schwarzschild mass $M_0$). Then the intersection between 
$\mathcal{S}_\ind{\emph{McV}}$ and $\mathcal{S}_\ind{\emph{cS}}$ 
is given by the subset of metrics in $\mathcal{S}_\ind{\emph{McV}}$ 
with constant $m$ or, equivalently, by the subset of metrics in 
$\mathcal{S}_\ind{\emph{cS}}$ whose conformal factor has a
gradient proportional to the Killing field 
$\bpartial/\bpartial T$ of the Schwarzschild metric 
(see~(\ref{eq:Schw-metric}) for notation). 
\end{pro}
Note that we excluded the `trivial' cases in which $m$ or $M_0$ 
(or both) vanish for the following reason: Comparing the expressions 
for the Weyl part of the MS energy of the two types of metrics 
(see~(\ref{eq:Omega-condi}) in Appendix~\ref{sec:Intersection-McV-cS}) 
it follows that $m$ vanishes iff $M_0$ does and this, in turn, leads 
to a metric conformally related to the Minkowski metric where 
the conformal factor depends only on time, that is, a FLRW metric. 
But such a spacetime, being homogeneous, is not of interest to us here. 

In particular, Proposition~\ref{pro:Intersection-McV-cS} implies that the 
metric of Sultana and Dyer~\cite{Sultana.Dyer:2005} are \emph{not} of 
type~(\ref{eq:McVittieAnsatz}), as suggested in Section\,IV\,A 
of~\cite{Faraoni.Jacques:2007} and allegedly shown in Section\,II 
of~\cite{Faraoni:2009} (cf.~our footnote~\ref{fn:SD-coord-change} at 
page~\pageref{fn:SD-coord-change}). 
This immediately follows from the observation that the conformal factor, 
expressed as function of the standard Schwarzschild coordinates that appear 
in~(\ref{eq:Schw-metric}), 
is given by $\Omega(T,R)=(T + 2M_0 \ln(R/2M_0-1))^2$ (compare Eqs.~(8) 
and (9) of~\cite{Sultana.Dyer:2005}), which also depends on $R$ and
hence does not satisfy the condition of 
Proposition~\ref{pro:Intersection-McV-cS}. We will have to say more 
about this at the beginning of Section\,\ref{sec:GeneralisationsMcVittie}
and in Section\,\ref{sec:NullFluid}.

\subsection{Spatial Ricci-isotropy}
\label{sec:SpatialIsotropy}
An important feature of any metric that is covered by the 
ansatz~(\ref{eq:McVittieAnsatz}) is, that its Einstein tensor 
is spatially isotropic in the following sense: `Spatially' refers 
to the directions orthogonal to $\bpartial/\bpartial t$ and 
`isotropy' to the condition that the  spatial restriction of 
the spacetime's Einstein tensor is proportional to the spatial 
restriction of the metric. Note that, since the spacetime's 
metric is time dependent, the spatial restriction of the spacetime's 
Einstein or Ricci tensor is \emph{not} the same as the Einstein or 
Ricci tensor of the spatial sections with their induced metrics. 
Hence the notion of spatial isotropy of the Einstein tensor used 
here is not the same as saying that the induced metric of the 
slices is an Einstein metric. 

Given that the Einstein tensor of~(\ref{eq:McVittieAnsatz}) is spatially 
isotropic in the sense used here, it is then obvious that Einstein's equation 
will impose a severe restriction upon the matter's energy-momentum tensor, 
saying that it, too, must be spatially isotropic. The degree of 
specialization implied by this will be discussed in more detail 
below. Here we only remark that this observation already answers 
in the negative a question addressed, and left open, in the last 
paragraph of \cite{Faraoni.etal:2008}, of whether 
(\ref{eq:McVittieAnsatz}) is the most general spherically
symmetric solution describing a black hole embedded in a spatially 
flat FLRW background: It clearly is not. 

In passing we make the 
obvious remark that, since $\Ein=\Ric-(1/2)\Scal\g$, where $\Ric$
denotes the Ricci tensor, the Einstein tensor is spatially isotropic 
iff the same holds for the Ricci tensor. For this reason we 
will from now on refer to \emph{spatial Ricci-isotropy} to denote
the feature in question. 

Now, a way to actually show spatial Ricci-isotropy is to compute 
the components of the Einstein tensor with respect to the 
orthonormal tetrad $\{\be_\mu\}_{\mu\in\{0,\cdots, 3\}}$ 
of~(\ref{eq:McVittieAnsatz}) defined by
\begin{equation}\label{eq:DefOrthnFrame}
  \be_\mu := 
  \norm{\bpartial/\bpartial x^\mu}^{-1}\,\bpartial/\bpartial x^\mu \,,
\end{equation}
where $\{x^\mu\}=\{t,r,\theta,\varphi\}$. Here, and henceforth, we write 
$\norm{\bv}:=\sqrt{\vert\bg(\bv,\bv)\vert}$. Note that $\be_0,\be_1$ are 
orthogonal to and $\be_2,\be_3$ tangent to the 2-spheres of constant 
radius~$r$. The non-vanishing independent components of the Einstein tensor 
with respect to the orthonormal basis~(\ref{eq:DefOrthnFrame}) are: 
\begin{subequations}\label{eq:McV-Ein-tensor}
\begin{alignat}{3}
  &\Ein(\be_0,\be_0) &&= 3 F^2 \,,
   \label{eq:McV-Ein-00}\\
  &\Ein(\be_0,\be_1) &&= 
   \tfrac{2}{R^2}\left(\tfrac{A}{B}\right)^2
   (a \, m)\!\dot{\phantom{I}}\! \,,
   \label{eq:McV-Ein-01}\\
  &\Ein(\be_i,\be_j) &&= -\left( 3F^2 + 2\tfrac{A}{B}\dot{F} \right)\delta_{ij}
  \,, \label{eq:McV-Ein-ij}
\end{alignat}
\end{subequations}
where an overdot denotes differentiation along $\bpartial/\bpartial t$. 
Before explaining the functions $A$, $B$, $R$, and $F$, note that the 
spatial isotropy of the Einstein tensor follows immediately 
from~(\ref{eq:McV-Ein-ij}), since $\Ein(\be_i,\be_j)\propto\delta_{ij}$. 
In~(\ref{eq:McV-Ein-tensor}) and in the following we set: 
\begin{equation}\label{eq:def-AB}
  A(t,r) := 1+m(t)/2r \,, \quad{  } B(t,r) := 1-m(t)/2r \,,
\end{equation}
and
\begin{equation}\label{eq:McVittieArealRadius}
  R(t,r) = \left( 1+\frac{m(t)}{2r} \right)^2 a(t)\,r \,,
\end{equation}
where $R$ is the areal radius for the McVittie 
ansatz~(\ref{eq:McVittieAnsatz}), and also
\begin{equation}\label{eq:def-F}
  F := \frac{\dot a}{a} + \frac{1}{rB}\frac{(a\,m)^{\bdot}}{a} \,.
\end{equation}
In passing we note that both quantities, $F$ and $a\,m$, that appear 
in the components of the Einstein tensor, have a geometrical 
interpretation: the 
former is one third the expansion of the vector field $\be_0$, that 
is, $F=\Div(\be_0)/3$, and the latter is the Weyl part of the 
Misner--Sharp energy of the metric~(\ref{eq:McVittieAnsatz}) 
(see~(\ref{eq:McV-MSE-Weyl}), below). Moreover, as we already noted 
in Section~\ref{sec:McVittieModel}, the observer field $\be_0$ is free 
of vorticity and shear. Hence, taking into account the 
relation~(\ref{eq:shear-scalar-expansion-rel}) between the expansion 
$\theta$ and the shear scalar $\sigma$ of an arbitrary spherically-symmetric 
observer field, the expansion of $\be_0$ can be simply written as 
$3\ed R(\be_0)/R$ so that $F$ may be expressed as 
\begin{equation}\label{eq:F-as-dR}
  F = \ed R(\be_0)/R \,.
\end{equation}

In order to estimate the degree of specialization implied by
spatial Ricci-isotropy, we ask for the most 
general spherically symmetric metric for which this is the 
case. To answer this, we first note that any spherically 
symmetric metric can always be written in the form  
\begin{equation}\label{eq:ss-metric-isotropic}
  \g=\left(\frac{B(t,r)}{A(t,r)}\right)^2\ed t^2 - 
  a^2(t) A^4(t,r) (\ed r^2 + r^2\gStwo)\,.
\end{equation}
This reduces to McVittie's ansatz~(\ref{eq:McVittieAnsatz}) if $A,B$ 
are given by~(\ref{eq:def-AB}). For the general spherically symmetric metric 
(\ref{eq:ss-metric-isotropic}), spatial Ricci-isotropy can be shown to be 
equivalent to 
\begin{equation}\label{eq:CondSpatialIsotropy} 
  \delta^2(AB) - 8(\delta A)(\delta B) = 0 \,,
\end{equation}
where $\delta:=r^{-1}\partial/\partial r=2\partial/\partial r^2$. 
It is obvious that there are many more solutions to this differential 
equation than just~(\ref{eq:def-AB}).

\subsection{Misner--Sharp energy}
\label{sec: MisnerSharpEnergy}
In order to be able to interpret~(\ref{eq:McVittieAnsatz}) as an ansatz for 
an inhomogeneity in a FLRW universe, it is useful to compute the Misner--Sharp 
(MS) energy and, in particular, its Ricci and Weyl parts. This concept of 
quasi-local mass, which is defined only for spherically symmetric spacetimes, 
and which in this case coincides with Hawking's more general 
definition~\cite{Hawking:1968} of quasi-local mass 
(see e.g.~\cite{Carrera.Giulini:2008a}), allows to detect localized 
sources of gravity. 

We recall the geometric definition of the MS 
energy~\cite{Misner.Sharp:1964,Hernandez.Misner:1966}:
\begin{equation}\label{eq:MS-energy-original-def}
  E := -\tfrac{1}{2} R^3 \,K\,,
\end{equation}
where $R$ denotes the areal radius and $K$ the extrinsic curvature.
More precisely, the equation should be read and understood as 
follows: First of all, the quantities $R$ and $K$, and hence also 
$E$, are real-valued functions on spacetime. In order to determine 
their values at a point $p$, recall that, due to the 
requirement of spherical symmetry, there is a unique 
two-(or zero-) dimensional $SO(3)$ orbit $S(p)$ through $p$. 
The value of $R$ at $p$ is as explained below 
Eq.~(\ref{eq:Hubble-law-McVittie}) and the value of 
$K$ at $p$ is 
\begin{equation}\label{eq:DefSecCurvature}
  K(p):=\frac{\Riem(\bX_p,\bY_p,\bX_p,\bY_p)}
        {\g(\bX_p,\bX_p)\g(\bY_p,\bY_p) - \bigl(\g(\bX_p,\bY_p)\bigr)^2} \,. 
\end{equation}
Here $\Riem$ is the (totally covariant) Riemannian curvature
tensor of spacetime and $\bX_p$ and $\bY_p$ are any two linearly 
independent vectors in the tangent space at $p$ which are also 
tangent to the orbit $S(p)$. Note that the right-hand side only
depends of the plane spanned by $\bX_p,\bY_p$ and not on the 
vectors spanning it. Finally we note that the minus sign 
in~(\ref{eq:MS-energy-original-def}) is just a relict of our 
signature choice (mostly minus). 

From the curvature decomposition for a spherically symmetric 
metric (see~\cite{Carrera.Giulini:2008a}) one can rewrite  
(\ref{eq:MS-energy-original-def}) in the form
\begin{equation}\label{eq:MS-energy}
  E = \frac{R}{2} \big( 1 + \g(\Grad R,\Grad R) \big) \,,
\end{equation}
where $\Grad R$ denotes the gradient vector-field of $R$. This 
provides a convenient expression for the computation of the MS 
energy. For a self-contained review of the basic properties of the MS energy 
as well as its interpretation as the amount of active gravitational energy 
contained in the interior of the spheres of symmetry ($SO(3)$-orbits) and 
its relation with the other mass concepts, see~\cite{Carrera.Giulini:2008a}. 

The decomposition of the Riemann tensor into a Ricci and a Weyl part 
leads, together with~(\ref{eq:MS-energy-original-def}), to a natural 
decomposition of the MS energy into a Ricci and Weyl part (see 
also~\cite{Carrera.Giulini:2008a}). For the Ricci part of the MS 
energy of~(\ref{eq:McVittieAnsatz}) we get
\begin{equation}\label{eq:McV-MSE-Ricci}
  E_\ind{R} = \tfrac{1}{6}R^3 \Ein(\be_0,\be_0) 
           = \frac{R}{2} (\ed R(\be_0))^2 \,.
\end{equation}
The first equality in~(\ref{eq:McV-MSE-Ricci}) can be derived 
by merely using the spatial Ricci-isotropy in the expression for 
the Ricci part of the Riemann tensor. The second equality follows 
then with~(\ref{eq:McV-Ein-00}) and~(\ref{eq:F-as-dR}). 
The Weyl part can now be obtained as the difference between the 
full MS energy and (\ref{eq:McV-MSE-Ricci}). We use the expression 
(\ref{eq:MS-energy}) for the former and write 
$\g(\Grad R,\Grad R)=\bigl(\be_0(R)\bigr)^2-\bigl(\be_1(R)\bigr)^2$. 
The part involving $\be_0(R)$ equals the Ricci part~(\ref{eq:McV-MSE-Ricci}) 
and hence the Weyl part is given by $(R/2)\bigl(1-(\be_1(R))^2\bigr)$. 
From~(\ref{eq:McVittieArealRadius}) we calculate $\be_1(R)$ and hence 
obtain for the Weyl part of the MS energy: 
\begin{equation}\label{eq:McV-MSE-Weyl}
  E_\ind{W} = a\,m\,.
\end{equation}
The Ricci part of the MS energy is that part which, via Einstein's 
equation, can be \emph{locally} related to the matter's energy-momentum 
tensor, whereas the gravitational mass of the central object is contained 
in the Weyl part of the MS energy. Notice that the latter is spatially 
constant (the functions $a$ and $m$ in (\ref{eq:McV-MSE-Weyl}) only
depend on time) but may depend on time. If the latter is the case we 
interpret this as saying that the central mass exchanges energy with 
the ambient matter.

\subsection{Singularities and trapped surfaces}
\label{sec:SingTrappedSurf}
Next we comment on the singularity properties of the McVittie 
ansatz~(\ref{eq:McVittieAnsatz}). From~(\ref{eq:McV-Ein-ij}) one suspect, 
because of the term proportional to $1/B$, a singularity in the Ricci part 
of the curvature at $r=m/2$ (that is at $R=2am=2E_\ind{W}$). In fact, this 
corresponds to a genuine curvature singularity, as one can see from 
looking, for example, at the following expression for the scalar curvature 
(i.e.~the Ricci scalar), 
\begin{equation}\label{eq:Scal-McV}
  \Scal = - 12F^2 - 6\tfrac{A}{B}\dot{F} \,, 
\end{equation}
which can be quickly computed from~(\ref{eq:McV-Ein-tensor}). 
In Appendix~\ref{sec:McV-Ricci-sing} we insert into this expression 
the definition~(\ref{eq:def-F}) of $F$ and expand this in powers of 
$1/(rB)$. This allows to prove 

\begin{pro}\label{pro:McV-Ricci-sing}
The Ricci scalar for a metric of the form (\ref{eq:McVittieAnsatz}) 
becomes singular in the limit $r \to m/2$ for any functions $a$ and 
$m$, except for the following three special cases: 
\begin{compactenum}[$(i)$]
\item $m=0$ and $a$ arbitrary (FLRW), 
\item $a$ and $m$ are constant (Schwarzschild), and 
\item $(a\,m)^{\bdot}\!=0$ and $(\dot a/a)^{\bdot}\!=0$ 
(Schwarzschild--de\,Sitter). 
\end{compactenum}
\end{pro}

\noindent
This means that, as long as we stick to the ansatz~(\ref{eq:McVittieAnsatz}), 
at $r=m/2$ there will always (with the only exceptions listed above) be a 
singularity in the Ricci part of the curvature and thus, assuming Einstein's 
equation is satisfied, also in the energy momentum tensor, irrespectively of 
the details of the underlying matter model. Hence any attempt to eliminate 
this singularity by maintaining the ansatz~(\ref{eq:McVittieAnsatz}) and 
merely modifying the matter model is doomed to fail. 

In particular, this is true for the generalizations presented in 
\cite{Faraoni.Jacques:2007}, contrary to what is claimed in that work 
and its follow ups~\cite{Gao.etal:2008,Faraoni.etal:2008}. We also 
remark that it makes no sense to absorb the singular factors $1/B$ 
in front of the time derivatives by writing $(A/B)\bpartial/\bpartial t$ 
as $\be_0$ and then argue, as was done in \cite{Faraoni.Jacques:2007}, 
that this eliminates the singularity. The point is simply that then 
$e_0$ applied to any continuously differentiable function diverges as 
$r\rightarrow m/2$. Below we will argue that this singularity lies 
within a trapped region. 

Specializing to the McVittie model, recall that in this case it is 
assumed that the fluid moves along the integral curves of 
$\bpartial/\bpartial t$, which become lightlike in the limit as $r$ 
tends to $m/2$. Their acceleration is given by the gradient of the 
pressure, which necessarily diverges in the limit $r\rightarrow m/2$, 
as one explicitly sees from~(\ref{eq:accel-McVittie}). 
For a more detailed study of the geometric 
singularity at $r=m/2$, see~\cite{Nolan:1999a,Nolan:1999b}. 

For spherically symmetric spacetimes the Weyl part of the curvature has 
only a single independent component, which is $-2/R^3$ times the Weyl 
part of the MS energy, by the very definition of the latter 
(see~\cite{Carrera.Giulini:2008a}). The square of the Weyl tensor 
for the ansatz~(\ref{eq:McVittieAnsatz}) may then be conveniently 
expressed as 
\begin{equation}\label{eq:Weyl-norm-McV}
  \sprod{\Weyl}{\Weyl} = 48\frac{(a\,m)^2}{R^6} \,. 
\end{equation}
This shows that $R=0$ also corresponds to a genuine curvature singularity, 
though this is not part of the region covered by our original coordinate 
system, for which $r>m/2$ (that is $R>2E_\ind{W}$).

It is instructive to also determine the trapped regions of McVittie 
spacetime. Recall that a spacelike 2-sphere $S$ is said to be 
\emph{trapped, marginally trapped}, or \emph{untrapped} if the product 
$\theta^+\theta^-$ of the expansions (for the definitions see 
e.g.~\cite{Carrera.Giulini:2008a}) for the ingoing and outgoing 
future-pointing null vector fields normal to $S$  is positive, zero, or 
negative, respectively. Taking $S$ to be $S_R$, that is, an $SO(3)$ 
orbit with areal radius $R$, it immediately follows from the 
relation $2\,\theta^+\theta^- = \g(\Grad R,\Grad R)/R^2$ 
(see~\cite{Carrera.Giulini:2008a}) that $S_R$ is trapped, marginally 
trapped, or untrapped iff $\g(\Grad R,\Grad R)$ is positive, zero, or 
negative, respectively. This corresponds to timelike, lightlike, or 
spacelike $\ed R$, or equivalently, in view of~(\ref{eq:MS-energy}), 
to $2E-R$ being positive, zero, or negative, respectively. 
Using~(\ref{eq:McV-MSE-Ricci}) together with~(\ref{eq:McV-Ein-00}), the 
MS energy for the McVittie ansatz can be written as 
$E = E_\ind{W} + R^3F^2/2$, so that 
\begin{equation}\label{eq:polynome-trapped}
  2E-R = F^2R^3 - R + R_S \,.
\end{equation}
Here we defined the `Schwarzschild radius' as $R_S:=2E_\ind{W}$, which  
generally will depend on time. We wish to determine the values of 
the radial coordinate ($r$ or $R$) at which the 
expression~(\ref{eq:polynome-trapped}) assumes the value zero. 
We shall continue to work with $R$ rather than $r$ since $R$ 
has the proper geometric meaning of areal radius. In the region 
we are considering (that is $r>m/2$ or, equivalently, $R>R_S$) 
the inversion of~(\ref{eq:McVittieArealRadius}) reads 
$r(R)=R \bigl( 1-R_S/2R+\sqrt{1-R_S/R} \bigr)/2a$, so 
that~(\ref{eq:polynome-trapped}) divided by $R_S$ can be 
written in the form 
\begin{equation}\label{eq:polynome-trapped-x}
  \frac{2E-R}{R_S} = 
  \left(\eta+\frac{\varepsilon}{x-1+\sqrt{x(x-1)}}\right)^2x^3 - x + 1 \,.
\end{equation}
Here we introduced the dimensionless radial coordinate $x:=R/R_S$ and 
the (small) parameters $\varepsilon:=\dot{R_S}$ and $\eta:=R_S/R_H$, 
where $R_H:=1/H$ denotes the `Hubble radius'. Recall that since $R>R_S$ 
we have $x>1$. 

Consider first the McVittie case, in which $\varepsilon=\dot R_S=0$. 
Then~(\ref{eq:polynome-trapped-x}) turns into a cubic polynomial in 
$x$ which is positive for $x=0$ and tends to $\pm\infty$ for 
$x\rightarrow\pm\infty$. Hence it always has a negative zero 
(which does not interest us) and two positive zeros iff 
\begin{equation}\label{eq:discrim-condition}
  R_S/R_H < 2/3\sqrt{3} \approx 0.38 \,.
\end{equation}
This clearly corresponds to the physical relevant case where the 
Schwarzschild radius is much smaller than the Hubble radius. One zero lies 
in the vicinity of the Schwarzschild radius and one in the vicinity of the 
Hubble radius, corresponding to two marginally trapped spheres. The exact 
expressions for the zeros can be easily written down, but are not very 
illuminating. In leading order in the small parameter $\eta=R_S/R_H$, they 
are approximated by 
\begin{subequations}\label{eq:McV-trapped}
\begin{align}
  &R_1 = R_S \bigl( 1 + \eta^2 + O(\eta^4) \bigr)\,, 
\label{eq:McV-trapped-1}\\
  &R_2 = R_H \bigl( 1 - \eta/2 + O(\eta^2) \bigr) \,.  
\label{eq:McV-trapped-2}
\end{align}
\end{subequations}
From this one sees that for the McVittie ansatz the radius of the marginally 
trapped sphere of Schwarzschild spacetime ($R_S$) increases and that of the 
FLRW spacetime ($R_H$) decreases. The first feature can, for the McVittie 
model, be understood as an effect of the cosmological environment, 
whereas the latter is an effect of the inhomogeneity in form of a 
central mass abundance. All the spheres with $R<R_1$ or $R>R_2$ are 
trapped and those with $R_1<R<R_2$ are untrapped. In particular, 
the singularity $r=m/2$, that is $R=2E_\ind{W}=R_S$, lies within the 
inner trapped region.

In the case in which $\varepsilon=\dot{R_S}$ is non-zero and `small' (see 
below in which sense), we expect that the zeros~(\ref{eq:McV-trapped}) vary 
smoothly in $\varepsilon$ so that, in particular, the singularity at $R=R_S$ 
still remains within the inner trapped region. An expansion in $\varepsilon$ 
gives, for the zero in the vicinity of the Schwarzschild radius: 
\begin{equation}\label{eq:McV-trapped-1-gen}
  R_1(\varepsilon) = R_S\bigl( 1+\eta^2 + (2-2\eta+13\eta^2)\varepsilon + 
                     O(\eta^3,\varepsilon^2) \bigr) \,,
\end{equation}
which clearly reduces to~(\ref{eq:McV-trapped-1}) for $\varepsilon=0$. 
From this expression one sees that, according to the physical expectation, 
in case of accretion ($\varepsilon>0$) the inner marginally trapped sphere 
becomes larger in area, whereas in the opposite case ($\varepsilon<0$) 
it shrinks. In our approximation (\ref{eq:McV-trapped-1-gen}), the 
singularity $R=R_S$ continues to lie inside the trapped region for 
`accretion rates' $\varepsilon=\dot{R_S}>-\eta^2/2$ or, in terms of 
physical quantities and re-introducing the factors of $c$, for 
$\dot{R_S}/c>-(R_S/R_H)^2/2$. However, this also characterizes the 
region of validity of the expansion~(\ref{eq:McV-trapped-1-gen}): 
Given a positive $\eta$, an expansion in $\varepsilon$ around zero 
exists only for $\varepsilon>-\eta^2/2$ since there exists no 
expansion on both $(\varepsilon,\eta)$ around $(0,0)$ (this is 
because the partial derivative of~(\ref{eq:polynome-trapped-x}) 
with respect to $x$ does not exist at $x=1$).

\subsection{Other global aspects}
\label{sec:OtherGlobAsp}
Another aspect concerns the global behavior of the McVittie 
ansatz~(\ref{eq:McVittieAnsatz}). We note that each hypersurface of constant 
time $t$ is a complete Riemannian manifold, which, besides the rotational 
symmetry, admits a discrete isometry given in $(r,\theta,\varphi)$ 
coordinates by 
\begin{equation}\label{eq:McVittieIsometry}
  \phi(r,\theta,\varphi) = 
  \bigl((m/2)^2\,r^{-1}\,,\,\theta\,,\,\varphi\bigr)\,.
\end{equation} 
This corresponds to an inversion at the 2-sphere $r=m/2$, which shows 
that the hypersurfaces of constant $t$ can be thought of as two isometric 
asymptotically-flat pieces joined together at the 2-sphere $r=m/2$.
This 2-sphere is totally geodesic since it is a fixed-point set of 
an isometry; in particular, it is a minimal surface.  
Except for the time-dependent factor $m(t)$, this is just like for the 
slices of constant Killing time in the Schwarzschild metric (the difference 
being that (\ref{eq:McVittieIsometry}) does not extend to an isometry of the 
spacetime metric unless $\dot m=0$). Now, the fact that $r\rightarrow 0$ 
corresponds to an asymptotically flat end of each of the 
3-manifolds $t=\mathrm{const.}$ implies that the McVittie metric 
cannot literally be interpreted as corresponding to a point particle 
sitting at $r=0$ ($r=0$ is in infinite metric distance) in an otherwise 
spatially flat FLRW universe, just like the Schwarzschild metric does not 
correspond to a point particle sitting at $r=0$ in Minkowski space. 
Unfortunately, McVittie seems to have interpreted his solution in this 
fashion~\cite{McVittie:1933} which even until recently gave rise to 
some confusion in the literature 
(e.g.~\cite{Gautreau:1984b,Sussman:1988,Ferraris.etal:1996}). 
A clarification was given in~\cite{Nolan:1999a}.

\section{Attempts to generalize McVittie's model}
\label{sec:GeneralisationsMcVittie}
The first obvious generalization consists in allowing for a 
non-vanishing cosmological constant. However, as was already 
indicated before, this is rather trivial since it merely 
corresponds to the substitutions $\varrho \to \varrho+\varrho_\Lambda$ 
and $p\to p+p_\Lambda$ in~(\ref{eq:McV-Einstein}), where 
$\varrho_\Lambda:=\Lambda/8\pi$ and $p_\Lambda:=-\Lambda/8\pi$ are the 
energy-density and pressure associated to the cosmological constant 
$\Lambda$. 

The attempts to non-trivially generalize the McVittie solution have 
focused so far on keeping the ansatz~(\ref{eq:McVittieAnsatz}) and 
relaxing the conditions on the matter in various ways. 
In \cite{Faraoni.Jacques:2007} generalization were presented allowing 
radial fluid motions relative to the observer vector field 
$\bpartial/\bpartial t$ (that is relaxing condition~(\ref{eq:u-McVittie})) 
as well as including heat conduction. Below we will critically review 
these attempts, taking due care of the geometric constraints 
imposed by the ansatz~(\ref{eq:McVittieAnsatz}), and also outline 
how to explicitly construct the respective solutions. 

Another exact solution that models an inhomogeneity in a cosmological 
spacetime was presented in~\cite{Sultana.Dyer:2005} by Sultana and Dyer 
and was recently analyzed in~\cite{Faraoni:2009}. Here the metric is 
conformally equivalent to the exterior Schwarzschild metric and the 
cosmological matter is composed of two non-interacting perfect fluids, 
one being pressureless dust, the other being a null fluid. One might ask 
if this solution fits into the class of McVittie models, as was suggested 
in~\cite{Faraoni.Jacques:2007}%
\footnote{\label{fn:McV-zero-pressure}In Section\,IV\,A 
of~\cite{Faraoni.Jacques:2007} it is suggested that the Sultana--Dyer 
metric is equal to the McVittie metric~(\ref{eq:McVittieAnsatz}) in which 
$a(t)=a_0 t^{2/3}$ and $m(t)=m_0$, for some constants $a_0$ and $m_0$ 
(see Eq.~(62) in~\cite{Faraoni.Jacques:2007}). Let denote the latter 
metric by $\tilde\g$ . Indeed, since $m$ is constant and in view of 
Proposition~\ref{pro:Intersection-McV-cS}, $\tilde\g$ is conformally 
related to the Schwarzschild metric. Moreover, as one may explicitly 
check via our Eq.~(\ref{eq:McV-Ein-ij}), the Einstein tensor of $\tilde\g$ 
has a vanishing spherical part. Despite sharing these two properties, 
$\tilde\g$ and the Sultana--Dyer metric are not equal. 
}
and allegedly confirmed explicitly in~\cite{Faraoni:2009}%
\footnote{\label{fn:SD-coord-change}The problem with the reasoning in 
Section\,II of \cite{Faraoni:2009} is the following (numbers refer to 
equations in \cite{Faraoni:2009}): It is true by construction that the 
Sultana--Dyer metric (2.1) is conformally related to the Schwarzschild 
metric, as expressed in the second line of (2.3) [the first line in (2.3) 
does not follow], but the conformal function $a$ depends non-trivially on 
the Schwarzschild coordinates for time \emph{and} radius (denoted by 
$\bar\eta$ and $\tilde r$ in \cite{Faraoni:2009}: Cf.~our discussion in 
the last paragraph of Section\,\ref{sec:ConfSchwClass}). Hence it is not 
possible to introduce a new time coordinate $\bar t$ that satisfies  
$d\bar t=ad\bar\eta$ (the right hand side is not a closed 1-form), as 
pretended in the transition to (2.5).}. 
However, as we already noted at the end of Section\,\ref{sec:ConfSchwClass} 
above in view of Proposition\,\ref{pro:Intersection-McV-cS}, this is not the 
case. Two further way to see this are as follows: First, the Sultana--Dyer 
metric is not spatially Ricci-isotropic%
\footnote{\label{fn:SD-not-spat-Ricci-iso}To show this, one has to show 
that there exists no timelike direction with respect to which the Ricci 
tensor (or, equivalently, the Einstein tensor) is spatially isotropic. 
This can be shown as follows: First note that the Einstein tensor of the 
Sultana--Dyer metric has the form 
$\Ein=\mu\buflat\otimes\buflat+\tau\bkflat\otimes\bkflat$ 
(see~\cite{Sultana.Dyer:2005}), where $\bu$ is a normalized future-pointing 
spherically-symmetric timelike vector field and $\bk$ the in-going 
future-pointing lightlike vector field orthogonal to the $SO(3)$-orbits 
normalized such that $\g(\bu,\bk)=1$. In particular, the spherical part 
of the Einstein tensor vanishes: Hence, the Einstein tensor is spatially 
isotropic iff there exists a non-vanishing spacelike spherically-symmetric 
(i.e.~orthogonal to the $SO(3)$-orbits) vector field $\bs$ with 
$\Ein(\bs,\bs)=0$. Without loss of generality one can chose $\bs$ to be 
normalized: $\bs=\sinh\chi\bu+\cosh\chi\be$, where $\be$ is the normalized 
vector field orthogonal to $\bu$ and to the $SO(3)$-orbits pointing in 
positive radial direction. Hence one has $\bk=\bu-\be$ and thus: 
$\Ein(\bs,\bs)=\mu\sinh^2(\chi)+\tau\exp(2\chi)$. Clearly, the latter 
expression vanishes nowhere in the physically interesting region 
(cf.~Eq.~(26) in~\cite{Sultana.Dyer:2005}), where both $\mu$ and $\tau$ 
are positive.} 
and, second, the McVittie metric is not compatible with the matter 
model used by Sultana and Dyer, with the sole exception of trivial or 
exotic cases, as will be shown in Section\,\ref{sec:NullFluid} below.

\subsection{Einstein's equation for the McVittie ansatz}
\label{sec:EinsteinEqMcV}
In the following we will restrict to those generalizations of the McVittie 
model which keep the metric ansatz~(\ref{eq:McVittieAnsatz}) and thus 
generalize only the matter model. For this purpose it is convenient to 
write down the Einstein's equation for an arbitrary spherically symmetric 
energy-momentum tensor $\bT$. Recall that spherical symmetry implies for 
the component of $\bT$ with respect to the orthonormal 
basis~(\ref{eq:DefOrthnFrame}) that $\bT(\be_a,\be_A)=0$ and 
$\bT(\be_A,\be_B)\propto\delta_{AB}$, where $a\in\{0,1\}$ and $A,B\in\{2,3\}$. 
Hence, the only independent, non-vanishing components of $\bT$ are 
\begin{subequations}\label{eq:T-ss}
\begin{alignat}{3}
  &S &&:= \bT(\be_0,\be_0)  \label{eq:def-S} \\
  &Q &&:= \bT(\be_1,\be_1)  \label{eq:def-Q} \\
  &P &&:= \bT(\be_2,\be_2)  \label{eq:def-P} \\
  &J &&:= -\bT(\be_0,\be_1) \label{eq:def-J} \,,
\end{alignat}
\end{subequations}
and these are functions which do not depend on the angular coordinates. 
Note that $S$ is the energy density, $Q$ and $P$ the radial and spherical 
pressure, and $J$ the energy flow---all referred to the observer field 
$\be_0$. The sing in~(\ref{eq:def-J}) is chosen such that a positive 
$J$ means a flow of energy in positive radial direction. 
Taking~(\ref{eq:T-ss}) into account, the Einstein equation for the McVittie 
ansatz~(\ref{eq:McVittieAnsatz}) and an arbitrary spherically symmetric 
energy-momentum tensor $\bT$ reduces to the following four equations: 
\begin{subequations}\label{eq:EinsteinEqMcVGen}
\begin{align}
  &(a\,m)^{\bdot} 
   = - 4\pi R^2 \left(\tfrac{B}{A}\right)^2 J \label{eq:EEMcVgen-1} \\
  &8\pi\, S = 3F^2                            \label{eq:EEMcVgen-2}\\
  &8\pi\, Q = - 3F^2 - 2\dot{F}^2\tfrac{A}{B} \label{eq:EEMcVgen-3}\\
  &P = Q                                      \label{eq:EEMcVgen-4} \,.
\end{align}
\end{subequations}
In view of~(\ref{eq:McV-MSE-Weyl}), the first equation relates the time 
variation of the Weyl part of the MS energy contained in the sphere of 
radius $R$ with the energy flow out of it. The last equation is nothing 
but spatial Ricci-isotropy. 

In the following subsections we will consider three models for the cosmological 
matter which generalize the original McVittie model: perfect fluid, perfect 
fluid plus heat flow, and perfect fluid plus null fluid.

\subsection{Perfect fluid}
\label{sec:RadialMotion}
Perhaps the simplest step one can take in trying to generalize the McVittie 
model is to stick to a single perfect fluid for the matter, but dropping the 
condition~(\ref{eq:u-McVittie}) of `no-infall' by allowing for radial motions 
relative to the $\bpartial/\bpartial t$ observer field. In this way one could 
hope to avoid a particular singular behavior in the pressure that may be due 
to the `no-infall' condition, though it is clear that the persisting geometric 
singularity must show up somehow in the matter variables as already discussed 
in Section~\ref{sec:SingTrappedSurf}. Unfortunately, as already shown 
in~\cite{Faraoni.Jacques:2007}, the relaxation of~(\ref{eq:u-McVittie}) 
does not lead to any new solutions. What we want to stress here is that 
the reason for this, as shown in more detail below, lies precisely in 
the restriction imposed by spatial Ricci-isotropy. 

We take thus the perfect-fluid energy-momentum tensor%
~(\ref{er:EMTensorForMcVittie}) for the matter and an arbitrary spherically 
symmetric four-velocity $\bu$. The latter is given in terms of the 
orthonormal basis for the metric~(\ref{eq:McVittieAnsatz}) by 
\begin{equation}\label{eq:u-McVittie-gen}
  \bu = \cosh\chi\,\be_0 + \sinh\chi\,\be_1 \,,
\end{equation}
where $\chi$ is the rapidity of $\bu$ with respect to the observer 
field $\be_0$ (a positive $\chi$ corresponds here to a boost in an 
outward-pointing radial direction). The non-vanishing components of 
the matter energy-momentum tensor~(\ref{er:EMTensorForMcVittie}) with 
four-velocity~(\ref{eq:u-McVittie-gen}) are:
\begin{subequations}\label{eq:McV-T-rapidity}
\begin{alignat}{3}
  &\bT(\be_0,\be_0) &&= \varrho + (\varrho+p)\sinh^2\chi \\
  &\bT(\be_0,\be_1) &&= - (\varrho+p)\sinh\chi\cosh\chi  \\
  &\bT(\be_1,\be_1) &&= p + (\varrho+p)\sinh^2\chi       \\
  &\bT(\be_2,\be_2) &&= \bT(\be_3,\be_3) = p\,. 
\end{alignat}
\end{subequations}

Clearly, the case of vanishing rapidity must lead to the original 
McVittie model. In this case, in fact, the matter energy-momentum 
tensor~(\ref{eq:McV-T-rapidity}) is already spatially isotropic 
so that~(\ref{eq:EEMcVgen-4}) is identically satisfied. Moreover, 
(\ref{eq:EEMcVgen-1}) implies $(a\,m)\!\dot{\phantom{I}}\! = 0$ and 
hence, in view of~(\ref{eq:def-F}), $F=\dot{a}/a$. Herewith Einstein's 
equation reduces to~(\ref{eq:McV-Einstein}) and thus one gets back the 
original McVittie model. 

In case of non-vanishing rapidity, spatial 
Ricci-isotropy~(\ref{eq:EEMcVgen-4}) implies the following constraint: 
\begin{equation}\label{eq:McV-constraint-1}
  \varrho+p = 0 \,.
\end{equation}
This means that the energy momentum tensor~(\ref{er:EMTensorForMcVittie}) 
has the form of a cosmological constant (using~(\ref{eq:McV-constraint-1}) 
in $\Div\bT=0$ it implies $\ed p=0$ and this, in turn, using 
again~(\ref{eq:McV-constraint-1}), implies $\ed \varrho=0$) so that this 
case reduces to the Schwarzschild--de\,Sitter solution and hence does not 
provide the physical generalization originally hoped for.

\subsection{Perfect fluid plus heat flow}
\label{sec:RadialMotionAndHeatFlow}
In a next step one may keep (\ref{eq:u-McVittie-gen}) and drop the 
condition that the fluid be perfect, in the sense of allowing for radial 
heat conduction. This is described by a spatial vector field $\bq$ that
represents the current density of heat, which here corresponds to 
the current density of energy in the rest frame of the fluid. 
Hence $\bq$ is everywhere orthogonal to $\bu$.%
\footnote{We note that the parametrization of the energy-momentum 
tensor given in~\cite{Faraoni.Jacques:2007} is manifestly different. 
Whereas we parametrized it in the usual fashion in terms of quantities 
(energy density, pressure, current density of heat) that refer to the 
fluid's rest system, the authors of \cite{Faraoni.Jacques:2007} also 
write down (\ref{er:EMTensorForMcVittie-gen}) (their Eq.~(79)), but 
with $\bq$ orthogonal to $\be_0$ (compare their Eq.~(93)) rather than 
$\bu$, which affects also the definition of $\varrho$. In fact, marking 
their quantities with a prime, their expression (79) is equivalent to 
our (\ref{er:EMTensorForMcVittie-gen}) iff $p=p'$, $q=q'\,\cosh\chi$, 
and $\varrho=\varrho'-2q'\,\sinh\chi$.} 
The fluid's energy momentum tensor then reads 
\begin{equation}\label{er:EMTensorForMcVittie-gen}
\bT= \varrho\,\ul\bu\otimes\ul\bu
   + p\,(\ul\bu\otimes\ul\bu - \g)        
   + \ul\bu\otimes\ul\bq 
   + \ul\bq\otimes\ul\bu \,.
\end{equation}
Taking~(\ref{eq:u-McVittie-gen}) as fluid velocity and imposing the heat 
flow-vector $\bq$ to be spherically symmetric, we have 
\begin{equation}\label{eq:heat-flow}
\bq = q\, \be 
   := q\,(\sinh\chi\,\be_0 + \cosh\chi\,\be_1) \,, 
\end{equation}
where $q$ is a function of $(t,r)$. Note that a positive $q$ 
corresponds to heat flowing in an outward-pointing radial 
direction. The independent non-vanishing components of the energy-momentum 
tensor are now as follows:
\begin{subequations}\label{eq:McV-T-rapidity-heat}
\begin{alignat}{3}
\label{eq:McV-T-rapidity-heat-a}
  &\bT(\be_0,\be_0) &&= \varrho + \tanh\chi\bigl( (\varrho+p)\tanh\chi + 
   2\,q \bigr) \\
\label{eq:McV-T-rapidity-heat-b}
  &\bT(\be_0,\be_1) &&= q - 
  \cosh^2\chi\bigl( (\varrho+p)\tanh\chi + 2\,q \bigr) \\
\label{eq:McV-T-rapidity-heat-c}
  &\bT(\be_1,\be_1) &&= p + \tfrac{1}{2}\sinh(2\chi)
   \bigl( (\varrho+p)\tanh\chi + 2q \bigr) \\
\label{eq:McV-T-rapidity-heat-d}
  &\bT(\be_2,\be_2) &&= \bT(\be_3,\be_3) = p \,.
\end{alignat}
\end{subequations}

Consider first the case of vanishing rapidity. Then the energy-momentum 
tensor is already spatially isotropic and Einstein's 
equation~(\ref{eq:EinsteinEqMcVGen}) reduces to 
\begin{subequations}\label{eq:McV-Einstein-heat-only}
\begin{alignat}{3}
  &(a\,m)\!\dot{\phantom{I}}\! &&= -4\pi R^2\,q\,\left(\tfrac{B}{A}\right)^2 
  \label{eq:McV-E-h-ma-dot}\\
  &8\pi\varrho &&= 3 F^2 
  \label{eq:McV-E-h-rho}\\
  &8\pi p   &&= - 3F^2 - 2\dot{F}\tfrac{A}{B} \,.
  \label{eq:McV-E-h-p}
\end{alignat}
\end{subequations}
These are three PDEs (though only time derivatives occur) for the five 
functions $a,m,\varrho,p$, and $q$ so that the 
system~(\ref{eq:McV-Einstein-heat-only}) is clearly under-determining. 
However, it is not possible to freely specify any 
two of these five functions and then determine the the other three 
via~(\ref{eq:McV-Einstein-heat-only}). For example, since the 
left-hand side of~(\ref{eq:McV-E-h-ma-dot}) depends only on $t$, 
the same must hold for the r.h.s., which implies that 
$q=f(t)/r^2(1-(m/2r)^2)^2$, where 
$f(t)=-(a\,m)\!\!\dot{\phantom{m}}\!\!/4\pi a^2$. In particular, 
the heat flow must fall-off as $1/r^2$. 

The easiest way to generate a solution in the case of zero rapidity is to 
specify the two functions $a(t)$ and $m(t)$, then let $A,B,R,F$ be 
determined by the 
definitions~(\ref{eq:def-AB},\ref{eq:McVittieArealRadius},\ref{eq:def-F}), 
and finally let the Einstein 
equations~(\ref{eq:McV-E-h-ma-dot},\ref{eq:McV-E-h-rho},\ref{eq:McV-E-h-p}) 
determine $q,\varrho$, and $p$, respectively. Notice that if we happen 
to specify $a$ and $m$ such that $a\,m$ is a constant, this immediately 
implies $q=0$ and $F=\dot{a}/a$, which leads to the standard McVittie 
solutions. From~(\ref{eq:McV-E-h-ma-dot}) the following is evident: 
if $q>0$ ($q<0$), that is for outwardly (inwardly) pointing heat flow, 
the Weyl part of the MS energy decreases (increases), as one would expect. 

Now we turn to the general case with non-vanishing rapidity: As it was the 
case for the perfect fluid in the previous subsection, the 
condition~(\ref{eq:EEMcVgen-4}) of spatial Ricci-isotropy implies a 
constraint on the matter: 
\begin{subequations}\label{eq:McV-Einstein-rapidity-and-heat}
\begin{equation}\label{eq:McV-E-r-h-constraint}
  (\varrho+p)\tanh\chi + 2\,q = 0\,. 
\end{equation}
Using this, the other components of the Einstein's equation reduces to: 
\begin{alignat}{3}
  &(a\,m)\!\dot{\phantom{I}}\! &&= +4\pi R^2\,q\, \left(\tfrac{B}{A}\right)^2 
  \label{eq:McV-E-r-h-ma-dot}\\
  &8\pi\varrho &&= 3 F^2 
  \label{eq:McV-E-r-h-rho}\\
  &8\pi p   &&= - 3F^2 - 2\dot{F}\tfrac{A}{B} \,.
  \label{eq:McV-E-r-h-p}
\end{alignat}
\end{subequations}
These are almost the same as in the case of vanishing rapidity 
(see~(\ref{eq:McV-Einstein-heat-only})), except for the opposite sign 
on the right-hand side of (\ref{eq:McV-E-r-h-ma-dot}).
This simply results from the fact that, according to 
(\ref{eq:McV-T-rapidity-heat-b}), $J=-\bT(\be_0,\be_1)=q$ for vanishing 
rapidity, whereas, due to the constraint~(\ref{eq:McV-E-r-h-constraint}), 
$J=-\bT(\be_0,\be_1)=-q$ for non-vanishing rapidity. This will be further 
interpreted below. Notice that for the equation of state $\varrho+p=0$ 
(cosmological term) (\ref{eq:McV-E-r-h-constraint}) implies $q=0$, thus 
leading once more to the Schwarzschild--de\,Sitter solution (see comment 
below Eq.~(\ref{eq:McV-constraint-1})). Henceforth we assume 
$\varrho+p \neq 0$, which implies that one can solve the 
constraint~(\ref{eq:McV-E-r-h-constraint}) for the rapidity:
\begin{equation}\label{eq:chi-in-terms-of-q}
  \tanh\chi = - \frac{2\,q}{\varrho+p} \,
\end{equation}
provided that $\abs{2q/(\varrho+p)}<1$.

The Einstein equation gives now four equations for the six functions 
$a,m,\varrho,p,q$, and $\chi$. As in the case of vanishing rapidity, 
this system is under-determining and it is not possible to 
freely specify any two of these six functions and then determine the 
the other four. In a similar fashion as before, the easiest way to 
generate a solution is to specify the two functions $a(t)$ and $m(t)$, 
to let then the 
definitions~(\ref{eq:def-AB},\ref{eq:McVittieArealRadius},\ref{eq:def-F})
determine $A,B,R,F$, and finally use the Einstein equations~%
(\ref{eq:McV-E-r-h-ma-dot},\ref{eq:McV-E-r-h-rho},\ref{eq:McV-E-r-h-p})
and~(\ref{eq:chi-in-terms-of-q}) to determine $q,\varrho,p$, and $\chi$, 
respectively. Again, choosing $a$ and $m$ such that their product is 
constant implies $q=0$ and $F=\dot{a}/a$, which leads to the standard 
McVittie solutions. 

In passing we remark that the condition $\varrho+p>0$ can be 
expressed geometrically in terms of the second time-derivative  
of the areal radius. Indeed, adding either (\ref{eq:McV-E-h-rho}) 
to (\ref{eq:McV-E-h-p}) or (\ref{eq:McV-E-r-h-rho}) to 
(\ref{eq:McV-E-r-h-p}) we obtain, taking into account 
$\be_0=(A/B)\bpartial/\bpartial t$ and (\ref{eq:F-as-dR}):
\begin{equation}\label{eq:RhoPlusP-Geometrically}
  4\pi(\varrho+p) = -\be_0\left(\frac{\be_0(R)}{R}\right) \,,
\end{equation}
which is positive iff the rate of change $\be_0(R)/R$ is a 
decreasing function along the integral lines of the 
observer $e_0$. In other words, $\varrho+p$ is 
positive iff $\ln(R)$ is a concave function on the 
worldline of the observer $e_0$, which is implied by, 
but not equivalent to, the function $R$ being concave. 

From~(\ref{eq:chi-in-terms-of-q}) and~(\ref{eq:McV-E-r-h-ma-dot}), and 
assuming $\varrho+p>0$, one sees the following: If $\chi>0$ ($\chi<0$), 
that is for an outwardly (inwardly) moving fluid with respect to $\be_0$, 
we have $q<0$ ($q>0$), that is an inwardly (outwardly) pointing heat flow, 
and the Weyl part of the MS energy decreases (increases). 
This means that the heat flow's contribution to the change of 
$E_\ind{W}$ never compensates that of the fluid motion, quite in 
accord with naive expectation. Below we show that for small rapidities 
the contribution due to the heat flow is minus one-half that of the 
cosmological matter. 

Let us now return to the sign-difference of the right-hand sides of
(\ref{eq:McV-E-h-ma-dot}) and~(\ref{eq:McV-E-r-h-ma-dot}). 
From~(\ref{eq:McV-T-rapidity-heat}) one infers that $J$ is the sum 
of the two contributions coming from the heat flow
\begin{equation}\label{eq:j-heat}
  J_h := q (1+2\sinh^2\chi)
\end{equation}
and from cosmological matter
\begin{equation}\label{eq:j-cosmo}
  J_m := (\varrho + p)\sinh\chi\cosh\chi \,,
\end{equation}
respectively. The constraint~(\ref{eq:McV-E-r-h-constraint}) can be 
written in the form 
\begin{equation}\label{eq:McV-constraint-J}
  2\cosh^2\chi J_h + (1+2\sinh^2\chi)J_m = 0, 
\end{equation}
which, for small rapidities $\chi$ (that is neglecting quadratic terms in
$\chi$), implies $2J_h+J_m \approx 0$. In this approximation the spatial 
energy-momentum flow due to heat is minus one-half that due to the 
cosmological matter. For the total flow this implies 
$J=J_m+J_h \approx J_m/2 \approx -J_h$. Now the sign difference 
between~(\ref{eq:McV-E-h-ma-dot}) and~(\ref{eq:McV-E-r-h-ma-dot}) 
is understood as follows: In case of vanishing rapidity one has 
$J_m=0$, $J_h=q$ and hence $J=q$ (leading to~(\ref{eq:McV-E-h-ma-dot})), 
whereas a short calculation reveals that in case of non-vanishing 
rapidity the constraint~(\ref{eq:McV-constraint-J}) implies
$J=J_m+J_h=-q$, leading thus to~(\ref{eq:McV-E-r-h-ma-dot}).

\subsection{Perfect fluid plus null fluid}
\label{sec:NullFluid}
The last tentative generalization we consider is taking for matter the 
incoherent sum (meaning that the respective energy-momentum tensors adds) 
of a perfect fluid (possibly with non-vanishing pressure) and a null fluid 
(eventually representing electromagnetic radiation). This clearly contains 
as special case the matter model considered by Sultana and 
Dyer~\cite{Sultana.Dyer:2005} in which the pressure vanishes. 
We already stressed in 
Section~\ref{sec:ConfSchwClass} that the metric ansatz 
of~\cite{Sultana.Dyer:2005} is different from~(\ref{eq:McVittieAnsatz}). 
Here we show that the matter model of~\cite{Sultana.Dyer:2005} is 
essentially incompatible with (\ref{eq:McVittieAnsatz}) except for 
trivial or exotic cases.  

The matter model consists of an ordinary perfect fluid and a null fluid 
(e.g.~electromagnetic radiation) without mutual interaction. Hence the 
matter's energy-momentum tensor is just the sum 
of~(\ref{er:EMTensorForMcVittie}) and 
\begin{equation}\label{eq:EMTensorNull}
  \bT^{\pm}_{\rm nf} = \lambda^2 \, \ul\bl^\pm\!\otimes\ul\bl^\pm \,, 
\end{equation}
where $\lambda$ is some non-negative function of $t$ and $r$ and $\bl^+$ and 
$\bl^-$ are, respectively, the outgoing and ingoing future-pointing null 
vector fields orthogonal to the spheres of constant radius $r$ partially 
normalized such that $\g(\bl^+,\bl^-)=1$. (It remains a freedom 
$\bl^\pm \mapsto \alpha^{\pm 1}\bl^\pm$, where $\alpha$ is a positive 
function). Without loss of generality we make use of this freedom and choose:
\begin{equation}\label{eq:def-l-pm}
  \bl^\pm = (\be_0 \pm \be_1)/\sqrt{2} \,,
\end{equation}
where $\be_0$ and $\be_1$ are the vectors of the orthonormal 
frame~(\ref{eq:DefOrthnFrame}). The components of the whole energy-momentum 
tensor with respect to this frame are then:
\begin{subequations}\label{eq:McV-T-null}
\begin{alignat}{3}
\label{eq:McV-T-null-a}
  &\bT(\be_0,\be_0) &&= \varrho + (\varrho+p)\sinh^2\chi + 
   \tfrac{1}{2}\lambda^2 \\
\label{eq:McV-T-null-b}
  &\bT(\be_0,\be_1) &&= - (\varrho+p)\sinh\chi\cosh\chi 
   \mp \tfrac{1}{2}\lambda^2 \\
\label{eq:McV-T-null-c}
  &\bT(\be_1,\be_1) &&= p + (\varrho+p)\sinh^2\chi + \tfrac{1}{2}\lambda^2 \\
\label{eq:McV-T-null-d}
  &\bT(\be_2,\be_2) &&= \bT(\be_3,\be_3) = p \,.
\end{alignat}
\end{subequations}
Here and below the upper (lower) sign corresponds to the 
outgoing (ingoing) null field. 

In the present case, the condition~(\ref{eq:EEMcVgen-4}) of 
spatial Ricci-isotropy is equivalent to the constraint:  
\begin{equation}\label{eq:constraint-null}
  (\varrho+p)\sinh^2\chi + \tfrac{1}{2}\lambda^2 = 0 \,.
\end{equation}
In the physically relevant case in which $\varrho+p>0$ this equation has 
only the trivial solution $\chi=0$ and $\lambda=0$, which leads to the 
original McVittie model. In the case $\varrho+p=0$ (\ref{eq:constraint-null}) 
implies $\lambda=0$, leading thus to the Schwarzschild--de\,Sitter spacetime 
(see comment below~(\ref{eq:McV-constraint-1})). Hence, a new solution is 
only possible if the matter is of an exotic type that satisfies 
$\varrho+p<0$, which either violates the weak energy-condition 
($\varrho>0$), or, less catastrophically, the dominant-energy 
condition ($\varrho>\vert p\vert$). In particular, for the matter 
model considered by Sultana and Dyer, one would need to violate the 
weak energy-condition.

\section{Conclusion}
\label{sec:Conclusion}
We conclude by commenting on the the main differences between these 
generalizations and the original McVittie model. First we stress once 
more that neither allowing for a nonzero rapidity nor a nonzero 
heat flow can eliminate the singularity at $r=m/2$ ($R=2am$) 
(as erroneously stated in~\cite{Faraoni.Jacques:2007}). 
The only substantial new feature of these generalizations is 
that the Weyl part of the MS energy $E_\ind{W}=a\,m$ is not 
constant anymore. In view of the fact that the combination  
\begin{equation}\label{eq:m-over-r-McV}
  m/r = A^2E_\ind{W}/R \approx E_\ind{W}/R
\end{equation}
contained in the McVittie ansatz gives the `Newtonian' part of the 
potential in the slow-motion and weak-field approximation 
(see~\cite{Carrera.Giulini:2008a}), we deduce that in order to get 
the geodesic equation for the generalized McVittie model, it suffices 
to substitute $m_0$ with $E_\ind{W}$ in the equation of motion derived 
in~\cite{Carrera.Giulini:2008a}. This means that the strength of the 
central attraction varies in time according to~(\ref{eq:EEMcVgen-1}), 
leading to an in- or out-spiraling of the orbits if $\ed E_\ind{W}(\be_0)>0$ 
or $\ed E_\ind{W}(\be_0)<0$, respectively. 

We identified the origin of why we could not vary the rapidity and 
the heat flow independently in the condition 
(\ref{eq:McV-E-r-h-constraint}) of spatial Ricci-isotropy, which 
is built into the ansatz (\ref{eq:McVittieAnsatz}). We saw that 
this geometric feature renders this ansatz special, so that it 
would be improper to call it a \emph{general} ansatz for spherical 
inhomogeneities in a flat FLRW universe. It remains to be seen 
whether useful generalizations exist which are captured by equally
simple ans\"atze.

\begin{acknowledgments}
D.G. acknowledges support from the Albert-Einstein-Institute in Golm
and the QUEST Excellence Cluster.
\end{acknowledgments}


\appendix

\section{Proof of Proposition~\ref{pro:Intersection-McV-cS}}
\label{sec:Intersection-McV-cS}
In this appendix we compute the intersection of the set 
$\mathcal{S}_\ind{McV}$ of metrics of type~(\ref{eq:McVittieAnsatz}), 
which we denote in the following by $\gMcV$, with the set 
$\mathcal{S}_\ind{cS}$ of metrics conformally related to 
an exterior Schwarzschild metric. Explicitly, the 
latter are of the form 
\begin{subequations}\label{eq:cS-metric}
\begin{equation}\label{eq:cS-metric-alone}
  \gcS:=\Omega^2\gSchw \,,
\end{equation}
where 
\begin{equation}\label{eq:Schw-metric}
  \gSchw = \left(1-\frac{2M_0}{R}\right)\ed T^2 - 
  \left(1-\frac{2M_0}{R}\right)^{-1}\!\!\!\!\ed R^2 - R^2\gStwo \,,
\end{equation}
\end{subequations}
denotes the Schwarzschild metric with mass $M_0$ in `standard' coordinates. 
The question is: for which functions $a$ and $m$ and, respectively, for 
which function $\Omega$ and parameter $M_0$ does the equation 
$\gMcV=\Omega^2\gSchw$ hold? Such an equality can be eventually established 
by finding a coordinate transformation, $\phi$ say, between the coordinates%
\footnote{The transformation between the angular variables is just the 
identity.} 
$(t,r)$ in~(\ref{eq:McVittieAnsatz}) and $(T,R)$ in~(\ref{eq:cS-metric}) which 
brings~(\ref{eq:McVittieAnsatz}) in form~(\ref{eq:cS-metric}). 
This involves solving coupled, non-linear partial differential equations 
for $\phi$, which depend on the four unknown parameter $a,m,\Omega$, and 
$M_0$. Needless to say that this is not really a thankful task. 
Alternatively, a better approach would be to compare all the independent, 
algebraic curvature-invariants of the two metrics: This would lead to a 
system of equations between scalars which involves the coordinate 
transformation $\phi$ in an algebraic way (i.e.~non differentiated). 

We adopt here an approach which is somewhere in the middle: First, we use 
just three invariants (the areal radius and the Ricci and the Weyl part of 
the MS energy) to drastically restrict the form of the coordinate 
transformation (see~(\ref{eq:coord-transf})) and derive thereby constraints 
on the free parameters $a,m,\Omega$, and $M_0$ (see~(\ref{eq:Omega-condi}) 
and~(\ref{eq:m-condi})). Second, we perform this restricted coordinate 
transformation and determine it completely. 
To simplify the calculation, instead of $\gMcV=\Omega^2\gSchw$, 
we consider the equivalent equation $\Omega^{-2}\gMcV=\gSchw$. In fact, for 
the Schwarzschild metric~(\ref{eq:Schw-metric}) it is immediate that the above 
mentioned quantities are, respectively, given by: 
\begin{alignat}{3}
  &R(\gSchw)         &&= R   \,, \label{eq:R-Schw} \\
  &E_\ind{W}(\gSchw) &&= M_0 \,, \label{eq:MS-W-Schw} \\
  &E_\ind{R}(\gSchw) &&= 0   \,. \label{eq:MS-R-Schw}
\end{alignat}
In order to compute the respective quantities for the metric 
$\Omega^{-2}\gMcV$ we first give their scaling behavior under conformal 
transformations. 

Clearly, because of their very definitions, for the areal radius and the 
Weyl part of the MS energy it holds:
\begin{equation}\label{eq:scaling-R}
  R(\Psi^2\g) = \Psi R(\g)
\end{equation}
and
\begin{equation}\label{eq:scaling-MS-Weyl}
  E_\ind{W}(\Psi^2\g) = \Psi E_\ind{W}(\g) \,, 
\end{equation}
respectively. For the whole MS energy it easily follows 
from~(\ref{eq:MS-energy}) and~(\ref{eq:scaling-R}):
\begin{equation}\label{eq:scaling-MS}
\begin{split}
  E(\Psi^2\g) = \Psi \Bigl( E(\g) + R^2 \g(\Grad R,\Grad\ln\Psi) \\ 
              + \tfrac{1}{2} R^3 \g(\Grad\ln\Psi,\Grad\ln\Psi) \Bigr) \,,
\end{split}
\end{equation}
where all the quantities on the r.h.s.~are referred to the metric $\g$. 
Hence, taking the difference between~(\ref{eq:scaling-MS}) 
and~(\ref{eq:scaling-MS-Weyl}) one gets that the Ricci part of the MS 
energy scales exactly like the whole MS energy, that is according 
to~(\ref{eq:scaling-MS}). 

Using these scaling properties together with~(\ref{eq:McVittieArealRadius}) 
and~(\ref{eq:McV-MSE-Weyl}) we get immediately: 
\begin{alignat}{3}
  &R(\Omega^{-2}\gMcV) &&= \Omega^{-1}(1+m/2r)^2ar     \label{eq:R-left}\\
  &E_\ind{W}(\Omega^{-2}\gMcV) &&= \Omega^{-1}a\,m \,. \label{eq:MS-W-left}
\end{alignat}
The equality between the Weyl part of the MS energy~(\ref{eq:MS-W-left}) 
and~(\ref{eq:MS-W-Schw}) implies 
\begin{equation}\label{eq:Omega-condi}
  \Omega(t,r)\,M_0 = a(t)\,m(t) \,,
\end{equation}
which gives a condition between the parameter $a,m,\Omega$, and $M_0$. 
Since we assumed that $M_0$ is positive, (\ref{eq:Omega-condi}) can be read 
as the expression for the conformal factor in the $(t,r)$ coordinates. This, 
together with the equality between the areal radius~(\ref{eq:R-left}) 
and~(\ref{eq:R-Schw}), implies in turn 
\begin{equation}\label{eq:transf-R}
  R(t,r) = \frac{M_0}{m(t)}(1+m(t)/2r)^2r \,, 
\end{equation}
which gives the first component of the coordinate transformation $\phi$. 
Now, using the scaling property~(\ref{eq:scaling-MS}) for the Ricci part 
of the MS energy, the expressions~(\ref{eq:McV-MSE-Ricci}) 
and~(\ref{eq:McVittieArealRadius}) for the Ricci part of the MS energy 
and, respectively, the areal radius of the McVittie metric ansatz, 
and~(\ref{eq:Omega-condi}) for the conformal factor, one gets, after 
some computations, 
\begin{equation}\label{eq:MS-R-left}
  E_\ind{R}(\Omega^{-2}\gMcV) = \frac{M_0}{2\,a\,m}(A^2ar)^3
    \left( \frac{\dot m}{m} \right)^2 \,.
\end{equation}
The equality between~(\ref{eq:MS-R-left}) and~(\ref{eq:MS-R-Schw}) then 
implies 
\begin{equation}\label{eq:m-condi}
  \dot m = 0 \,,
\end{equation}
that is $m = m_0$ for some positive constant $m_0$. This, in turns, implies 
that the transformation~(\ref{eq:transf-R}) for $R$ depends only on $r$ and 
not on $t$. Since the metrics are both in diagonal form, this implies that 
the transformation for $T$ must depend on $t$ only. 

Summarizing, so far we have seen that a set of necessary conditions for the 
equality of the two metrics implies the constraints~(\ref{eq:Omega-condi}) 
and~(\ref{eq:m-condi}) and that the coordinate transformation between 
$(t,r)$ and $(T,R)$ is of the form 
\begin{subequations}\label{eq:coord-transf}
\begin{alignat}{3}
  &T(t) &&=f(t)                               \label{eq:coord-transf-T} \\
  &R(r) &&=\tfrac{M_0}{m_0}(1+m_0/2r)^2 r \,, \label{eq:coord-transf-R}
\end{alignat}
\end{subequations}
for some differentiable function $f$ of $t$. Now, explicitly expressing 
the metric $\Omega^2\gSchw$ in the $(t,r)$ coordinates according 
to the coordinate transformation~(\ref{eq:coord-transf}) and the 
constraints~(\ref{eq:Omega-condi}) and~(\ref{eq:m-condi}), and putting 
the result equal to $\gMcV$, the only new condition that one gets is 
\begin{equation}\label{eq:f-condi}
  \dot f = \pm \frac{M_0}{m_0}\,\frac{1}{a} \,. 
\end{equation}
Here, the plus can be chosen in order to exclude a time inversion. 
It is important to note that~(\ref{eq:f-condi}) (together with an initial 
value) determines $f$ uniquely and do not give any constraint on the 
parameters $a,m,\Omega$, and $M_0$: The only constraints remain 
thus~(\ref{eq:Omega-condi}) and~(\ref{eq:m-condi}). 

The proof is concluded noticing that~(\ref{eq:Omega-condi}) means that the 
only constraint on $\Omega$ is that, expressed in the $(t,r)$ coordinates, 
it depends on $t$ only and hence, in view of~(\ref{eq:coord-transf-T}) and 
expressed in the $(T,R)$ coordinates, that it depends on $T$ only. More 
geometrically, this can be restated saying that the gradient of $\Omega$ 
must be proportional to $\bpartial/\bpartial T$, the Killing field of the 
Schwarzschild metric (see~(\ref{eq:Schw-metric})).

\section{Proof of Proposition~\ref{pro:McV-Ricci-sing}}
\label{sec:McV-Ricci-sing}
Inserting the definition~(\ref{eq:def-F}) of $F$ in the 
expression~(\ref{eq:Scal-McV}) for the Ricci scalar and organizing the 
result in powers of $rB \equiv r-m/2$ we get: 
\begin{equation}\label{eq:ScalExplicit}
\begin{split}
  \Scal=&-12\left(\frac{\dot a}{a}\right)^2 \\
        &-\frac{6}{rB}\left( 4\frac{\dot{a}(a\,m)\!\dot{\phantom{i}}}{a^2} 
         + rA\left(\frac{\dot a}{a}\right)\!\!\!\!\dot{\phantom{\frac{I}{I}}} 
         \right) \\
        &-\frac{6}{(rB)^2}\left( 
            2\left(\frac{(a\,m)\!\dot{\phantom{i}}}{a}\right)^2 
            \!\!+\!rA\!\left(\frac{(a\,m)\!\ddot{\phantom{i}}}{a}
                 -\frac{\dot{a}(a\,m)\!\dot{\phantom{i}}}{a^2}\right) 
           \right) \\
        &-\frac{3}{(rB)^3}rA\frac{\dot{m}(a\,m)\!\dot{\phantom{i}}}{a} \,.
\end{split}
\end{equation}
Hence, the Ricci scalar remains finite in the limit $r \to m/2$ iff all 
the three coefficient of $(rB)^{-k}$, for $k\in\{1,2,3\}$, vanish in this 
limit, that is iff it holds: 
\begin{subequations}\label{eq:ScalFinite}
\begin{align}
  &4\frac{\dot{a}(a\,m)\!\dot{\phantom{i}}}{a^2} 
   + m\left(\frac{\dot a}{a}\right)\!\!\!\!\dot{\phantom{\frac{I}{I}}} = 0 
  \label{eq:ScalFinite-1}\\
  &2\left(\frac{(a\,m)\!\dot{\phantom{i}}}{a}\right)^2 
            \!+\!m\left(\frac{(a\,m)\!\ddot{\phantom{i}}}{a}
                 -\frac{\dot{a}(a\,m)\!\dot{\phantom{i}}}{a^2}\right) = 0 
  \label{eq:ScalFinite-2}\\
  &\dot{m}(a\,m)\!\dot{\phantom{i}} = 0 
  \label{eq:ScalFinite-3}\,.
\end{align}
\end{subequations}
These conditions are clearly understood to hold for all times $t$ in which 
the functions $a$ and $m$ and their derivative exist. 
In view of~(\ref{eq:ScalFinite-3}) we have to distinguish between two cases: 
$\dot m=0$ and $(a\,m)^{\bdot}=0$, respectively. 
In the first case the system~(\ref{eq:ScalFinite}) reduces to the set of 
conditions 
$m(\ddot{a}/a+3(\dot{a}/a)^2)=0$ and 
$m(\ddot{a}/a+(\dot{a}/a)^2)=0$, 
which, in turn, reduces to $m=0$ (and $a$ arbitrary), corresponding to the 
FLRW metric, or to $\dot a=0$ (and $\dot m=0$), corresponding to the 
Schwarzschild metric. 
In the second case, in which $(a\,m)^{\bdot}=0$, (\ref{eq:ScalFinite}) 
reduces to $m(\dot{a}/a)^{\bdot}=0$, which implies either $m=0$ (and $a$ 
arbitrary), corresponding again to the FLRW metric, or 
$(\dot{a}/a)^{\bdot}=0$. Together with $(a\,m)^{\bdot}=0$, the latter 
corresponds to a McVittie metric with exponentially-growing (or -falling) 
scale factor $a(t)$, that is to a Schwarzschild--de\,Sitter metric.

\section{Shear-free observer fields in spherically symmetric spacetimes}
\label{sec:ProofShearFree}
Towards the end of Section\,\ref{sec:McVittieModel} we made use 
of the following result: A spherically symmetric normalized timelike 
vector field $\bu$ in a spherically symmetric spacetime $(\M,\g)$
is shear free iff the metric $\bh_\bu$ that $\g$ induces on the 
subbundle $\bu^\perp:=\{\bv\in T\M\mid \bg(\bv,\bu)=0\}$ by 
restriction is conformally flat. 

To prove this, we first note that the subbundle $\bu^\perp$ is
integrable, in other words, $\bu$ is hypersurface orthogonal. 
This follows from the spherical symmetry of $\bu$, 
which implies that $\bu^\perp$ contains the vectors tangent to 
the 2-dimensional $SO(3)$ orbits. Hence $\bu$ essentially lives in the 
2-dimensional orbit space%
\footnote{The orbit space is the quotient $\M/\!\sim$, where $\sim$ is 
the equivalence relation whose equivalence classes are the orbits. It is 
a manifold on the subset corresponding to 2-sphere orbits, to which we 
restrict attention here. To say that ``$\bu$ essentially lives in the 
orbit space'' means that \ul{$\bu$}~is the pull-back of a 1-form on the 
quotient via the natural projection.}, where it is trivially hypersurface 
orthogonal. The hypersurfaces orthogonal to $\bu$ in 
4-dimensional spacetime are then the preimages under the natural 
projection of the hypersurfaces (curves) in the 2-dimensional orbit space.   

As a result, we may now locally introduce so-called 
isochronous comoving coordinates,  with respect to which 
$\bu=A(t,r)^{-1}\bpartial/\bpartial t$ and 
\begin{equation}\label{eq:SphSymMetShear-1}
  \g =
  A^2(t,r)\,\ed t^2 - 
  B^2(t,r)\,\ed r^2 -
  R^2(t,r)\, \gStwo \,.
\end{equation}
(Note the different meanings of the functions $A$ and $B$ as compared to 
(\ref{eq:ss-metric-isotropic})). We now consider the tangent-space 
endomorphisms $\bnabla\bu : \bX\mapsto\bnabla_\bX\bu$ and their 
projection into the orthogonal complement of $\bu$, i.e., 
\begin{equation}\label{eq:SpatialEndomorphism}
  \bnabla^\perp \bu:= 
  \bigl(\bP^\perp_\bu\circ\bnabla\bu\circ\bP^\perp_\bu\bigr)\big\vert_{\bu^\perp}\,,
\end{equation}
where $\bP^\perp_\bu := \Id - \bu\otimes\ul\bu$ is the projection orthogonal 
to $\bu$ ($\Id$ is the identity endomorphism in the tangent spaces of $\M$). 
Note that $\bnabla^\perp \bu$ is symmetric due to the hypersurface 
orthogonality of $\bu$. A direct computation using 
(\ref{eq:SphSymMetShear-1}) yields
\begin{equation}\label{eq:SphSymMetShear-2}
  \bnabla^\perp\bu = 
  \bu\bigl(\ln(B)\bigr)\,\bP_r + 
  \bu\bigl(\ln(R)\bigr)\,\bP_\Stwo \,,
\end{equation}
where $\bP_r$ and $\bP_\Stwo$ are the projections parallel to 
$\bpartial/\bpartial r$ and parallel to the tangent 2-planes to the 
$S^2$-orbits, respectively. The trace $\theta$ of $\bnabla^\perp\bu$, 
which gives the expansion of $\bu$, is 
$\theta = \bu\bigl(\ln(B)\bigr) + 2\bu\bigl(\ln(R)\bigr)$, so that the 
trace-free part of $\bnabla^\perp\bu$, known as the shear endomorphism 
$\bsigma$, is given by: 
\begin{equation}\label{eq:SphSymMetShear-3}
  \bsigma := \nabla^\perp\bu - \tfrac{1}{3}\theta\,\Id^\perp
           = \sigma\,(\bP_\Stwo - 2\bP_r) \,, 
\end{equation}
where $\sigma:=\tfrac{1}{3}\,\bu\bigl(\ln(R/B)\bigr)$ denotes the shear scalar 
(only defined in a spherically-symmetric setting) and 
$\Id^\perp = \bP_r + \bP_\Stwo$ the identity endomorphism in $\bu^\perp$. 
In passing, we note that the defining equations for $\theta$ and 
$\sigma$ just given immediately lead to the following simple 
relation between the shear scalar, expansion, and the variation 
of the areal radius along~$\bu$, that we made use of in 
Section\,\ref{sec:SpatialIsotropy}: 
\begin{equation}\label{eq:shear-scalar-expansion-rel}
  \sigma + \theta/3 = \bu(\ln(R)) \,.
\end{equation}
Now, according to (\ref{eq:SphSymMetShear-3}), the shear of 
$\bu$ vanishes iff the shear scalar $\sigma$ does, that is, 
iff $\bu\bigl(\ln(R/B)\bigr)$ vanishes. This is equivalent to 
$R/B$ being independent of $t$ or to $R(t,r)=\mu(r)B(t,r)$
for some function $\mu$, so that the line element 
(\ref{eq:SphSymMetShear-1}) can be rewritten in the spatially 
conformally flat form 
\begin{subequations}
\begin{equation}\label{eq:SphSymMetShear-4}
  \g = 
  \tilde{A}^2(t,\rho)\,\ed t^2 - 
  \tilde{C}^2(t,\rho)\bigl(\ed\rho^2+\rho^2\,\gStwo\bigr) \,, 
\end{equation}
where $\tilde{A}(t,\rho):=A(t,r(\rho))$, $\tilde{C}(t,\rho):=C(t,r(\rho))$, 
and 
\begin{equation}\label{eq:SphSymMetShear-5}
  C(t,r)=\frac{B(t,r)\mu(r)}{\rho(r)}
\end{equation}
with
\begin{equation}\label{eq:SphSymMetShear-6}
  \rho(r) = \rho_0\exp\left\{ \int_{r_0}^{r}\frac{dr'}{\mu(r')} \right\} \,.
\end{equation}
\end{subequations}
Hence we see that vanishing shear of $\bu$ implies conformal flatness of 
the corresponding spatial metric. For the converse we first note that, 
since $\bu$ and $\g$ are spherically symmetric, the spatial metric 
$\bh_\bu$ is itself spherically symmetric, so that $\g$ can be written in the 
form~(\ref{eq:SphSymMetShear-4}). This implies that the corresponding $R/B$ 
depends only on the radial coordinate and hence that the shear of $\bu$ 
vanishes.


\bibliographystyle{apsrmplong-spires}
\bibliography{COSMOLOGY}

\end{document}